\documentclass{article}

\usepackage{interact}

\usepackage{setspace}
\usepackage{color}
\usepackage{graphicx}
\usepackage{adjustbox}
\usepackage{hyperref}
\usepackage[binary-units]{siunitx}
\usepackage{tabularx}
\usepackage{enumitem}
\usepackage[normalem]{ulem}
\usepackage{multirow}
\usepackage{caption}
\usepackage[super,comma]{natbib}

\usepackage[acronym,shortcuts]{glossaries}
\newacronym{BLV}{BLV}{blindness or low vision}
\newacronym{CCTV}{CCTV}{closed circuit television}
\newacronym{HCI}{HCI}{human-computer interaction}
\newacronym{HMD}{HMD}{head-mounted display}
\newacronym{AMD}{AMD}{age-related macular degeneration}
\newacronym{AR}{AR}{augmented reality}
\newacronym{FOV}{FOV}{field of view}
\newacronym{IADL}{IADL}{instrumental activity of daily living}
\newacronym{LV}{LV}{low vision} 
\newacronym{MR}{MR}{mixed reality}
\newacronym{PLV}{PLV}{people with low vision}
\newacronym{PRISMA}{PRISMA}{preferred reporting items for systematic reviews and meta-analyses}
\newacronym{PRL}{PRL}{preferred retinal locus}
\newacronym{RP}{RP}{retinitis pigmentosa}
\newacronym{SLV}{SLV}{simulated low vision}
\newacronym{SPV}{SPV}{simulated prosthetic vision}
\newacronym{VFL}{VFL}{visual field loss}
\newacronym{VR}{VR}{virtual reality}
\newacronym{XR}{XR}{extended reality}
\newacronym{IoT}{IoT}{Internet of Things}
\glsdisablehyper  

\begin{document}


\title{A Systematic Review of Extended Reality (XR) for Understanding and Augmenting Vision Loss}

\author{
    \name{Justin Kasowski\textsuperscript{1,*}\thanks{CONTACT Justin Kasowski. Email: justin\_kasowski@ucsb.edu}, Byron A. Johnson\textsuperscript{2,*}, Ryan Neydavood\textsuperscript{2}, Anvitha Akkaraju\textsuperscript{2}, and Michael Beyeler\textsuperscript{2,3}}
    \affil{
        \textsuperscript{1}Graduate Program in Dynamical Neuroscience, University of California, Santa Barbara, CA, 93106 \\
        \textsuperscript{2}Department of Psychological \& Brain Sciences, University of California, Santa Barbara, CA, 93106 \\
        \textsuperscript{3}Department of Computer Science, University of California, Santa Barbara, CA, 93106 \\
        \textsuperscript{*}These authors contributed equally
    }
}

\maketitle

\newpage
\begin{abstract}
\emph{Purpose:} Over the past decade, extended reality (XR) has emerged as an assistive technology not only to augment residual vision of people losing their sight but also to study the rudimentary vision restored to blind people by a visual neuroprosthesis. To make the best use of these emerging technologies, it is valuable and timely to understand the state of this research and identify any shortcomings that are present.
\emph{Methods:} Here we present a systematic literature review of 227 publications from 106 different venues assessing the potential of XR technology to further visual accessibility. In contrast to other reviews, we sample studies from multiple scientific disciplines, focus on augmentation of a person's residual vision, and require studies to feature a quantitative evaluation with appropriate end users.
\emph{Results:} We summarize prominent findings from different XR research areas, show how the landscape has changed over the last decade, and identify scientific gaps in the literature.
Specifically, we highlight the need for real-world validation, the broadening of end-user participation, and a more nuanced understanding of the suitability and usability of different XR-based accessibility aids.
\emph{Translational Relevance:} By broadening end-user participation to early stages of the design process and shifting the focus from behavioral performance to qualitative assessments of usability, future research has the potential to develop XR technologies that may not only allow for studying vision loss, but also enable novel visual accessibility aids with the potential to impact the lives of millions of people living with vision loss.
\end{abstract}


\newpage
\section{Introduction}

In recent years, rapid technological advances have led to an increase in the number of assistive technology and electronic mobility aids for people with \ac{BLV}
\cite{butler_technology_2021,manjari_survey_2020,brule_review_2020,htike_ability_2020}.
These assistive devices use various sensors (e.g., cameras, depth and ultrasonic sensors) to capture the environment and often apply computer vision and signal processing techniques to detect, recognize, or enhance text, people, and obstacles. 
Some systems translate vision into alternative modalities, such as converting a visual object into auditory or haptic signals \cite{manjari_survey_2020}.
Although vision substitution techniques would be essential for people who are completely blind, the majority of people with visual impairment have useful residual vision and prefer to use it to observe the environment \cite{htike_ability_2020}. Additionally, some of those who have lost their vision are eligible to receive a visual neuroprosthesis, which is a device that electronically stimulates neurons in the visual pathway to restore a rudimentary form of vision.

State-of-the-art devices facilitate the implementation of complex computer vision algorithms to drastically improve the functionality of residual vision. Specifically, \ac{XR} is a rapidly advancing technology with direct contributions to our understanding of low vision and an inherent ability to augment residual vision in a useful way. 
\Ac{XR} is a universal term that encompasses \ac{VR}, \ac{AR}, and other immersive \ac{MR} environments \cite{kardong-edgren_call_2019}. While there is some overlap among these terms, experiments with \ac{VR} technology generally allow researchers to monitor participants in a controlled 3D environment, whereas \ac{AR} devices integrate and enhance real-world surroundings.

To make the best use of these emerging technologies, it is valuable and timely to understand the state of this research and identify any shortcomings that are present.
This includes identifying visual enhancements and information processing techniques that have already been successful at improving visual function and quality of life for their end users. 
It is also valuable to take into account human factor considerations, such as the individual preferences and accessibility needs of people with different levels of residual vision and underlying conditions.
Accommodating these individual differences may determine the success and usability of \ac{XR}-based visual aids in the near future.

In serve of this, we make three contributions:
\begin{itemize}[topsep=0pt, noitemsep]
    \item We provide a systematic literature review of 227 publications from 106 different venues assessing the potential of \ac{XR} technology development to further visual accessibility.
    In contrast to other recent reviews \cite{manjari_survey_2020,machado_systematic_2021,htike_ability_2020}, we sample studies from multiple scientific disciplines, focus on accessibility aids that visually augment a person's residual vision, and require studies to feature a quantitative evaluation with appropriate end users.
    \item We perform a meta-analysis that gives a holistic view of the prevalent types of \ac{XR} technologies, approaches, and tasks used in \ac{BLV} research, and shows how the landscape has changed over the last decade.
    \item We summarize prominent findings from different \ac{XR} research areas and identify scientific gaps in the literature. Specifically, we highlight the need for real-world validation, the broadening of end-user participation, and a more nuanced understanding of the suitability and usability of different \ac{XR}-based accessibility aids.
\end{itemize}

\section{Related Work}

A number of review articles have reported on the potential and limitations of \ac{XR} technologies for people with visual impairment.
These reviews highlighted a multitude of sensor-based technologies, ranging from smartphones~\cite{manjari_survey_2020} to \ac{VR} \acp{HMD}~\cite{htike_ability_2020,aydindogan_applications_2021}, which could be used to recognize commercial products~\cite{machado_systematic_2021}, detect obstacles and reduce navigation time~\cite{santos_systematic_2021,htike_ability_2020}, or support social interactions~\cite{qiu_review_2022}.

However, these articles also pointed to several gaps in the literature and suggested potential avenues for future research.
On the technology side, some studies suggested to use smart clothing~\cite{santos_systematic_2021} for nearby obstacle detection and to integrate devices with existing \ac{IoT} infrastructure~\cite{machado_systematic_2021}.
On the behavioral side, Kelly and Smith~\cite{kelly_impact_2011} lamented that most studies in their review lacked methodological rigor.
More recently, Brule \emph{et al.}~\cite{brule_review_2020} highlighted the need for adequate quantitative empirical evaluation by involving \ac{BLV} end users in the design process.
This sentiment was shared by Butler \emph{et al.}~\cite{butler_technology_2021}, who further highlighted the need to broaden application areas and ask for more \emph{in situ} evaluation.
These works demonstrate that systematic reviews can highlight key issues in accessibility research and provoke the field to reflect and improve on their practice.

However, all of the above reviews focused mostly on technology that offered nonvisual feedback to the user (e.g., via text-to-speech software or vibrotactile devices), such as electronic travel aids, electronic orientation aids, position locator devices, and sensory substitution devices.
Although such devices can form the basis of practical accessibility aids, 
no systematic review has yet focused on \ac{XR} technology that uses \emph{vision} as the primary feedback mechanism; for instance, by visually augmenting the scene in an \ac{AR} display to enhance the end user's residual vision.

Given the rapid rate at which new research is being conducted in the use of these new assistive technologies, we consider it timely to review the field in order to better understand the nature and potential of this research.


\section{Methods}

\subsection{Scope}
\label{sec:scope}

The goal of this review is thus to summarize recent research in \ac{XR} applications for people with \ac{BLV} and identify trends that can inform the development of future assistive technologies.
This includes quantifying the number of studies, summarizing the major findings, identifying gaps in current practices, and making a number of specific recommendations for future research.
Specifically, the goal was to answer a number of questions regarding the use of \ac{XR} in \ac{BLV} research:
\begin{itemize}
    \item What are the main types of \ac{XR} technologies used in \ac{BLV} research?
    \item What experimental tasks are studied, and how?
    \item What are key challenges or scientific gaps that researchers should focus on in the future?
\end{itemize}

To define ``reality'' in \acf{XR}, we required that all studies had their end users interact with an immersive environment. Being able to move one's body has been identified as a key factor for immersion~\cite{pasch_immersion_2009}; hence we excluded studies that did not update the environment based on the participant's eye, head, or body movements.
Studies with monitors usually do not allow for head or body movements, but can still be used to represent a virtual space if complemented with head or eye tracking. 
Smart devices (and their applications) were only included if they were used to augment the environment in some way (as opposed to, for instance, augmenting a website).
For example, a text magnifier that could be pointed at text in real life would be included, because the first-person view would update based on what the magnifier is pointing at; but an on-screen text magnifying app would not.

While much has been written about the theoretical and technical aspects of accessibility technology, we specifically wanted to focus on studies that incorporated appropriate quantitative empirical evaluation, as suggested by \cite{brule_review_2020,butler_technology_2021}.
Articles were excluded if not an original work (e.g., review papers), if they solely proposed new technology or methods, or if they were focused on a survey about basic device use (i.e., ``how often do you use your smart device to read text'').
Survey studies were included if they focused on participants' perceived experience while using a specific technology.

\subsection{Systematic Review Process}

\begin{figure}[h]
    \centering
     \includegraphics[width=.8\textwidth]{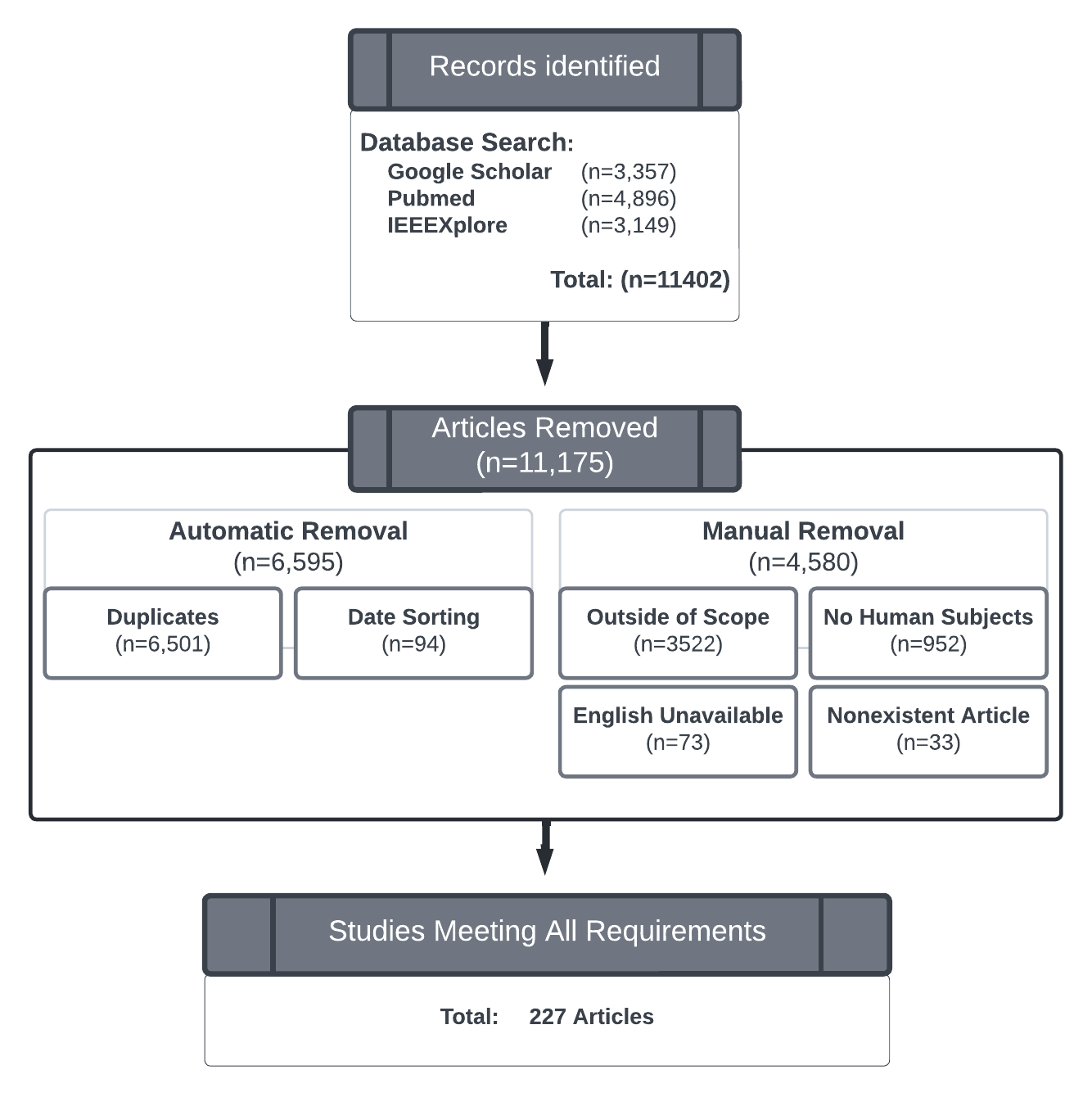} 
    \caption{\ PRISMA flow diagram. The results from three databases (Google Scholar, IEEE Xplore, and PubMed) were searched to identify work that combined \ac{XR} technology with low vision research. After removing duplicates, improperly dated studies, and studies that did not involve human subjects research, we ended up with 227 articles to be included in the review.}
    \label{fig:methods-prisma} 
\end{figure}

In contrast to a traditional review, systematic reviews can provide a more complete and less biased picture of the type of work being undertaking in the field and point to key challenges moving forward \cite{mulrow_systematic_1994}.
To help reduce bias and encourage a holistic review, we followed the PRISMA protocol~\cite{page_prisma_2021}, which is a method for systematically searching databases with a list of keywords and documenting every step (Figure~\ref{fig:methods-prisma}).
This includes reporting the number of papers excluded from further analysis along with the reasons for exclusion.

To cover a large body of research independent of their publication venue, we searched three databases (Google Scholar, IEEE Xplore, and PubMed) on the 17th of January, 2022. Each search included different keyword pairs (Table~\ref{tab:methods-keywords}) designed to identify work that combines \ac{XR} technology with low vision and accessibility research.
Each database was searched with all allowable search parameters that did not result in a full-text search; that is, we searched the title alone with Google Scholar, title/abstract in PubMed, and title/abstract/author keywords in IEEE Xplore.
This resulted in $11,402$ matches across the three databases.

\begin{table}[h]
\centering
    \begin{tabular}{cc|ccc}
        \hline
        \multicolumn{2}{c}{\bf{Visual Impairment }} & \multicolumn{3}{c}{\bf{Extended Reality}} \\
        \hline
``bionic vision''&``low*vision``& ``AR''& ``augment*''&``device*'' \\

``prosthetic vision'' & ``retinal implant'' &  ``display*'' & ``enhance*'' & ``head-mounted''\\

``retinal prosthesis'' & ``vis* aid*'' &``immersive'' & ``mixed'' & ``reality'' \\
``vis* loss'' & ``vis* impair*'' & ``simulat*''       & ``technolog*'' & ``wearable'' \\ \\

    \end{tabular}
    \caption{\ Keyword Combinations: Search terms used on Google Scholar, IEEE Xplore, and PubMed. Every ``Visual Impairment" term was combined with all ``Extended Reality" terms. ``*'' denotes the wildcard character.}
    \label{tab:methods-keywords}
\end{table}

Due to the nature of searching multiple databases with numerous keyword combinations, a large number of duplicate articles were identified. All articles were imported into Zotero, which identified $6,501$ duplicates and $94$ other articles whose publication date preceded the year 2010.

The remaining $4,807$ articles were reviewed by the research team and assessed for eligibility. A total of $4,580$ papers were manually removed.
The majority of these ($n=3,522$) were deemed outside the scope of the review as they presented a visual accessibility prototype that (even though it may operate on vision as an input modality) offered only nonvisual feedback to the user (see Section~\ref{sec:scope}).
In addition, we had to exclude a large number of theoretical studies that did not evaluate their proposed design on appropriate end users ($n=952$).
Furthermore, we removed $73$ papers not available in English and $33$ papers that could not be found online (most of these turned out to be manually entered citations on Google Scholar).

The remaining $227$ studies were included in the review.

\subsection{Interactive Collection}

The identified articles are available to the reader both as a BIB file (see Supplementary Materials) and as an interactive collection that can be accessed at \url{https://app.litmaps.com/shared/map/CE0C5D29-8F18-4F2D-9866-0BE1EA4AF288}.
Created with the free online platform ``Litmaps'', visitors are able to inspect individual articles and see how they are connected to other articles in the collection.

An example visualization is shown in Fig.~\ref{fig:cluster}, where each paper is represented by a circle whose size is proportional to the number of citations the paper received to date.
The publication date increases moving left to right, and papers are spread over the y-axis according to how similar their titles are. 
To calculate title similarity, Litmaps uses Allen AI's SPECTER model~\cite{cohan_specter_2020}, which projects the title of each paper into a 600-dimensional space before it is reduced to one dimension using UMAP.
This view can be customized at the above URL, allowing visitors to cluster by keyword, title similarity, or citation count.

\section{Meta-Analysis}

\subsection{Publications by Venue}

Upon completion of the systematic review, we identified 227 papers from 105 different venues.
57 of these were conference publications (Table~\ref{tab:venues-conferences}) that included full papers, workshop papers, and extended abstracts/short papers.
The majority of papers were classified as full papers by their respective venue.
Of the 35 different conference venues in our dataset, the most popular conference was the Annual Meeting of the IEEE Engineering in Medicine and Biology Society (EMBC) followed by ACM ASSETS and CHI.
Not surprisingly, the range of conference venues also included top-tier conferences in mixed reality (e.g., IEEE VR, ISMAR, and UIST), accessible technology (ICCHP, UAHCI), and ubiquitous computing (e.g., UbiComp, PerCom, ACIIW).
There were also a number of publications that were considered short papers or extended abstracts that accompanied a poster presentation or demo.
A small fraction of papers were part of a workshop or satellite event instead of the main conference track.

\begin{table}[h]
      \centering
        \begin{tabular}{l|rrr|r}
        Venue & Full & Short & Workshop & Total \\
        \hline
        IEEE EMBC & 11 & 0 & 0 & 11 \\
        ACM ASSETS & 4 & 2 & 0 & 6 \\
        ACM CHI & 4 & 1 & 0 & 5 \\
        ICCHP & 2 & 0 & 0 & 2 \\
        IEEE VR & 0 & 0 & 2 & 2 \\
        CVPR & 0 & 0 & 2 & 2 \\
        All others & 21 & 6 & 2 & 29 \\
        \hline
        {\bf Total} & {\bf 42} & {\bf 9} & {\bf 6} & {\bf 57}
        \end{tabular}
        \caption{\ Conference publications by venue}
        \label{tab:venues-conferences}
\end{table}

The other 170 publications were full-length articles that appeared in one of 70 different scientific journals (Table~\ref{tab:venues-journals}).
Here, the largest body of work appeared in vision science journals that specialize in either basic (e.g., Journal of Vision, Vision Research) or clinical research (e.g., Optometry \& Vision Science, Investigative Ophthalmology \& Visual Science, Translational Vision Science \& Technology).
A number of papers also appeared in biomedical engineering journals (e.g., Journal of Neural Engineering, IEEE Transactions) and general-purpose journals (e.g., PLOS ONE, Scientific Reports).

\begin{table}[h]
      \centering
        \begin{tabular}{lr}
        Venue & Count \\
        \hline
    
        Optometry \& Vision Science & 13 \\
        Vision Research & 13 \\
        Journal of Neural Engineering & 12 \\
        Journal of Vision & 12 \\
        Ophthalmic \& Physiological Optics (OPO) & 11 \\
        Artificial Organs & 8 \\
        PLOS ONE & 8 \\
        Translational Vision Science \& Technology (TVST) & 8 \\
        Investigative Ophthalmology \& Visual Science (IOVS) & 6 \\
        All others & 79 \\
        \hline
        {\bf Total} & {\bf 170}
        \end{tabular}
        \caption{\ Journal publications by venue}
        \label{tab:venues-journals}
\end{table}

\subsection{Publications by Research Area}

\begin{figure}[h]
    \centering
     \includegraphics[width=.93\textwidth]{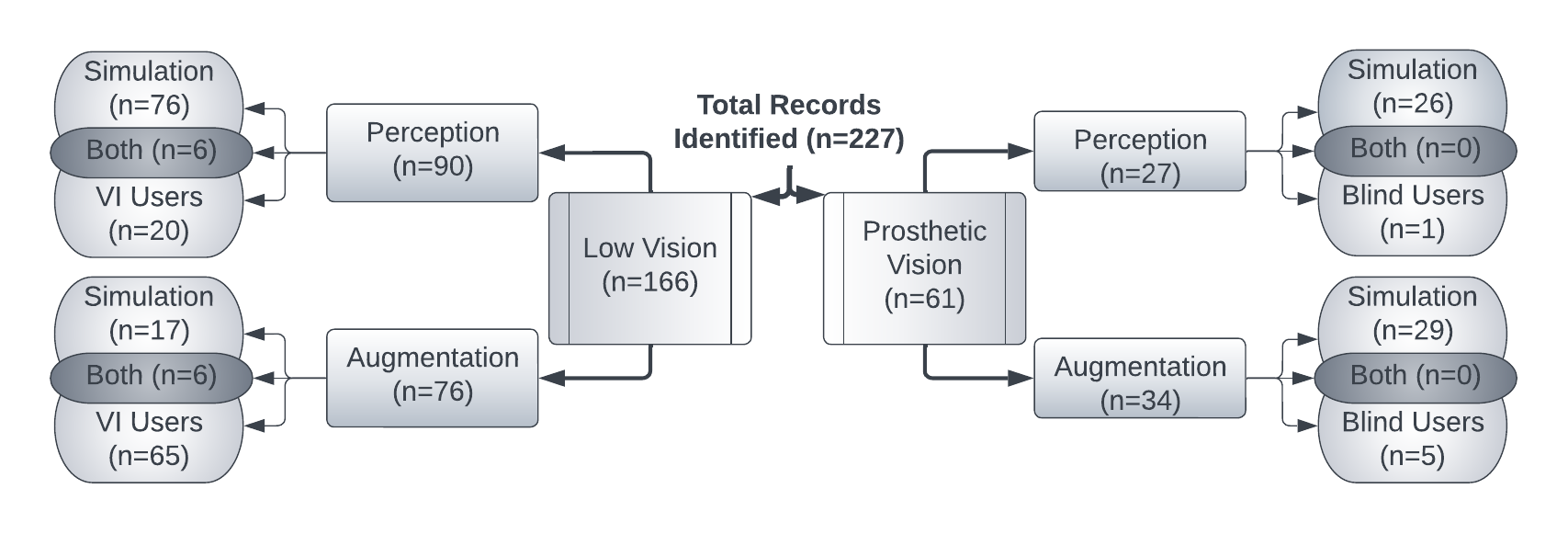} 
    \caption{\ The 227 articles included in this review were manually assessed and categorized by 1) whether the end users were people with low vision (defined as having some residual light perception) or people who were totally blind (no light perception), 2) whether the article used XR technology to study visual perception and behavior or proposed a new XR augmentation technology, and 3) whether the article involved BLV end users, simulations of the relevant impairment condition, or both.}
    \label{fig:groupings} 
\end{figure}

To get a better understanding of the research areas and applications covered by the corpus of identified papers, we inspected all 227 articles and thematically sorted them into one of four nested categories (Fig.~\ref{fig:groupings}):
\begin{itemize}[topsep=0pt]
    \item XR studies for people with low vision ($n=166$), defined as having some residual light perception, further divided into:
    \begin{itemize}
        \item ``perception`` studies where XR was used as a tool to study visual perception and behavior of people with low vision ($n=90$), often involving lenses, prisms, and \acp{HMD} to simulate a specific eye condition; and
        \item ``augmentation'' studies that primarily focused on the development of novel XR-based assistive devices and augmentation strategies ($n=76$). This included techniques like depth encoding, obstacle highlighting, and magnifications.
    \end{itemize}
    \item XR studies for people who are totally blind ($n=61$), defined as having no residual light perception, further divided into:
    \begin{itemize}
        \item ``perception'' studies where XR was used as a tool to study the visual perception and behavior of blind people whose vision was restored with a visual prosthesis ($n=27$), often studying the perceptual or behavioral consequences of a specific prosthesis design; and
        \item ``augmentation'' studies that used XR to augment the artificial vision provided by a visual prosthesis ($n=34$), often used to evaluate the effectiveness of specific visual augmentation strategies.
    \end{itemize}
\end{itemize}

As is evident from Fig.~\ref{fig:groupings}, 73\% of studies focused on low vision as opposed to blindness; however, within these two broad categories there was a roughly equal focus on augmentation and perception.
Interestingly, low vision augmentation studies extensively involved low vision participants (87\% of studies), whereas all three other categories predominantly relied on computer simulations of the visual condition under study that would be presented to sighted participants.
Roughly 7\% of low vision studies included both sighted participants (e.g., to evaluate a prototype using simulated low vision) and \ac{PLV} (e.g., to validate their system on appropriate end users).
This is in stark contrast to the prosthetic vision studies, none of which involved both sighted and blind participants.

To get a better understanding of the main types of XR technologies and experimental tasks used in BLV research (see research questions in Section~\ref{sec:scope}), we screened every article in the collection to identify which XR display type was used, which experimental task was studied (and how), and whether \ac{BLV} end users were involved (Table~\ref{tab:metrics}).

\begin{table}[h]
\centering
    \begin{tabular}{llr|rrrr|r}
        &&& \multicolumn{2}{c}{\textbf{Low}} & \multicolumn{2}{c|}{\textbf{Prosthetic}} &  \\
        &&& \multicolumn{2}{c}{\textbf{Vision}} & \multicolumn{2}{c|}{\textbf{Vision}} & \textbf{Total}  \\
         &&& \hspace{0.5cm} A & P & \hspace{0.5cm} A & P & \\
        \hline

        \textbf{Experimental task:}
        &\multicolumn{2}{r|}{Visual function testing} & 13 & 14 & 5 & 3 & 35\\
        &\multicolumn{2}{r|}{Visual search/recognition} & 37 & 42 & 19 & 16 & 114\\
        &\multicolumn{2}{r|}{Spatial cognition} & 26 & 34 & 10 & 8 & 78\\
         &  &  &  & & & \\

        \textbf{\ac{BLV} end user involvement:} &\multicolumn{2}{r|}
         {Pre-study qualitative assessment} & 30 & 16 &  0  & 0 & 46 \\
         &\multicolumn{2}{r|}{Evaluated BLV performance} & 52 & 18 & 5  & 1 & 76 \\
         &\multicolumn{2}{r|}{Post-study qualitative assessment} & 20 & 8 & 2 & 1 & 31 \\
        &  &  &  & & & \\
        
        \textbf{XR display type:} &\multicolumn{2}{r|}{Monitors} & 10 & 31 & 3 & 8 & 52 \\
        &\multicolumn{2}{r|}{Handheld devices} & 14 & 0 & 0 & 0 & 14\\
        &\multicolumn{2}{r|}{Nonelectronic wearables} & 8 & 39 & 0 & 0 & 47\\
        &\multicolumn{2}{r|}{VR wearables} & 3 & 16 & 20 & 16 & 55\\
        &\multicolumn{2}{r|}{AR wearables} & 41 & 4 & 11 & 3 & 59\\
         &  &  &  & & & \\
    \end{tabular}
    \caption{\ Experimental tasks studied, extent of \ac{BLV} end-user involvement, and XR display type used. Note that publications involve end users in multiple ways. If more than one task was studied, or more than one display type used, the more rigorous metric was used. A: augmentation, P: perception.}
    \label{tab:metrics}
\end{table}

All studies could be categorized as focusing either on low-level visual function measurements ($n=47$) such as acuity, contrast detection threshold, and orientation discrimination (e.g.,~\cite{tatiyosyan_standalone_2020,lene_changes_2020,butt_simulation_2015,almutleb_simulation_2018}), mid to high-level visual function tasks ($n=111$) such as visual search and object recognition (e.g.,~\cite{walsh_adaptation_2014,geringswald_central_2015, geringswald_impairment_2016, liu_integrating_2016}), or high-level spatial cognition tasks ($n=72$) such as wayfinding and obstacle avoidance tasks (e.g.,~\cite{alberti_visual_2014,zult_effects_2019,rand_spatial_2015,murray_simulated_2014}).

We were also interested in knowing whether these studies were conducted with input or feedback from \ac{BLV} end users by reporting at least one of the following:
\begin{itemize}[topsep=0pt]
    \item conduction of a pre-study qualitative assessment (e.g., surveys, questionnaires, or interviews) with \ac{BLV} participants by the study authors, which was used to inform the design of a device/application;
    \item evaluation of perceptual or behavioral performance of the proposed simulation, device, or application with \ac{BLV} end users;
    \item conduction of a post-study qualitative assessment (e.g., surveys or interviews) with \ac{BLV} participants by the study authors, which was used to report about the usability of a device/application.
\end{itemize}
These numbers are summarized in Table~\ref{tab:metrics}. While 76 studies~($33\%$) used \ac{BLV} end users to evaluate performance, most prosthetic vision studies ($95\%$) did not. Of the six studies that recruited bionic eye users, none consulted with \ac{BLV} users about their information needs, and only two studies based their work on previous findings about the information needs of prosthesis users~\cite{sadeghi_glow_2021, rachitskaya_computer-assisted_2020}. Additionally, while many studies used \ac{BLV} participants, very few~($n=31$) conducted post-study qualitative assessments. 

In terms of device types, \ac{VR} wearables were the most popular device used ($n=75$), followed by desktop monitors ($n=50$), nonelectronic wearables ($n=46$) and \ac{AR} wearables ($n=44$).
While all of these device types have been used in low vision research, prosthetic vision studies have so far been restricted to monitors and VR/AR wearables.

\subsection{Publications by Year}

Additionally, we wanted to know how the field has progressed over the last decade. Fig.~\ref{tab:year-area} summarizes the number of studies for each of our four major groups.
We found that the overall number of papers in the corpus has been steadily increasing over the last decade, largely driven by increasing interest in \ac{XR} for low vision (augmentation: 500.0\% increase, perception: 141.7\% increase since 2010).

\begin{table}[h]
    \centering
    \resizebox{\textwidth}{!}{\begin{tabular}{l|rrrrrr|r}
        Year                         & 2010-11   & 2012-13   & 2014-5    & 2016-17   & 2018-19   & 2020-21 & {\bf Total} \\
        \hline
        Low vision: & & & & & & & \\
        - XR for studying perception \& behavior                & 12        & 10        & 11        & 12        & 16        & 29   & {\bf 90} \\
        - XR for augmenting low vision & 4       & 4        & 8        & 15       & 21        & 24   & {\bf 76} \\
        Blindness: & & & & & & & \\
        - XR for studying perception \& behavior         & 5        & 2        & 8       & 5         & 4        & 3    & {\bf 27} \\
        - XR for augmenting prosthetic vision         & 4        & 6        & 6       & 5         & 7       & 6    & {\bf 34} \\
        \hline
        {\bf Total}                  & {\bf 25}  & {\bf 22}  & {\bf 33}  & {\bf 37}  & {\bf 48}  & {\bf 62} & {\bf 227}
    \end{tabular}}
    \caption{Number of publications per year and application area.}
    \label{tab:year-area}
\end{table}

Interestingly, the recent rise in popularity of \ac{XR} technologies for people with \ac{BLV} can be primarily attributed to \ac{VR} and \ac{AR} wearables (Table~\ref{tab:year-devices}), despite an increase in availability of handheld devices such as smartphones and tablets.
Research in \ac{VR} wearables increased by 112.5\% from 2010--11 to 2020--21, and monitor-based \ac{XR} research increased by 433.3\%.

\begin{table}[h]
    \begin{tabular}{l|rrrrrr|r}
         & 2010-11 & 2012-13 & 2014-15 & 2016-17 & 2018-19 & 2020-21 & Total \\
        \hline
        Monitors & 3 & 7 & 8 & 9 & 9 & 16 & {\bf 52} \\
        Handheld devices & 0 & 1 & 1 & 4 & 4 & 4 & {\bf 14} \\
        Nonelectronic wearables & 12 & 6 & 7 & 6 & 11 & 5 & {\bf 47} \\
        VR wearables &8 &5 &9 &7 &9 &17
        & {\bf 55} \\
        AR wearables &2 &3 &8 &11 &15 &20 & {\bf 59} \\
        \hline
        {\bf Total} & {\bf 25} & {\bf 22} & {\bf 33} & {\bf 37} & {\bf 48} & {\bf 62} & {\bf 227}
    \end{tabular}
    \caption{\ Number of publications per year and device type.}
    \label{tab:year-devices}
\end{table}

\section{Research Areas}
\label{sec:research-areas}

\Ac{XR} is a rapidly advancing technology that may not only enable near-future visual accessibility aids but may also serve as a tool to study visual perception and behavior of people with blindness or low vision.
Activity in this multidisciplinary space is not just confined to human-computer interaction, but also includes research in cognitive psychology and visual neuroscience as well as clinical applications.

In this section, we summarize the main research activities and findings across the four nested categories introduced in the previous section.
We also aim to identify trends that can inform the development of future assistive technologies.

\subsection{XR for Studying Perception \& Behavior of People with Low Vision}
\label{sec:XR-VI-behavior}

XR technology is increasingly being used to study visual perception and behavior of \acf{PLV} (39.6\% of papers in our collection).
Visual impairments such as \ac{AMD}, glaucoma, and \ac{RP} produce scotomas; that is, area(s) of the retina where the functioning of retinal cells is altered or diminished \cite{jones_seeing_2020,pollmann_intact_2020}.
Scotomas can lead to changes in visual function such as \ac{VFL}, which may affect perceptual or behavioral performance \cite{jones_seeing_2020,pollmann_intact_2020}.
Most of the studies in this category attempted to measure visual function either by recruiting \ac{PLV} for testing a specific task (e.g.,~\cite{miura_virtual_2018,h_hoppe_clevr_2020}) or by using low-vision simulations with sighted participants (e.g.,~\cite{jones_seeing_2020,seitz_we_2020}).

\subsubsection{XR for Simulating the Perception of People with Low Vision}

79 of the 90 identified low-vision perception studies (87.7\%) relied on \ac{SLV}.
An inexpensive means to simulate low vision for a sighted participant is the use of nonelectronic wearables, such as specially designed glasses, goggles, filters, and more (e.g.,~\cite{kobashi_effect_2012,morris_effects_2012,scott_perception_2012, kanzler_inertial_2016,latham_best_2011}).
Modern alternatives include desktop displays or \acp{HMD} that update the view based on where the user is looking (``gaze-contingent display''), which can be used to simulate specific eye conditions in real time (e.g.,~\cite{kwon_rapid_2013,seitz_we_2020,jones_seeing_2020}; also known as ``altered reality''~\cite{bao_augmented_2019}).
While \ac{VR} and \ac{AR} headsets allow for similar experimental designs, researchers have direct control of the environment when using \ac{VR}.
The primary advantage of this approach is the ability to flexibly remove, add, or modify many different features of visual input.

Simulations are a valuable experimental tool for studying performance in tasks such as visual search \cite{addleman_simulated_2020,jones_seeing_2020}, face perception \cite{liu_integrating_2016,tsank_domain_2017}, reading \cite{huang_augmented_2019,latham_best_2011}, and navigation \cite{barhorst-cates_navigating_2019,freedman_gaze_2019,zult_effects_2019}. 
A prime example of this is OpenVisSim~\cite{jones_seeing_2020}, which can simulate different impairments in real time (Fig.~\ref{fig:slv}).
To demonstrate the utility of OpenVisSim, Jones and colleagues~\cite{jones_seeing_2020} simulated a central scotoma in \ac{VR} (based on perimetric data from a person with glaucoma) and had sighted participants ($n=23$) perform a visual search task with a Fove 0 \ac{HMD} and a mobility task with the HTC Vive.
They demonstrated that the scotoma led to impaired performance in both tasks and found that a scotoma located in the upper visual field (inferior retina) led to worse performance, more eye movements, and more head movements.

However, the most commonly studied topic concerned the consequences of \ac{VFL} on eye movements and associated behavior.
It is well-known that people with central \ac{VFL} shift their oculomotor reference location from the fovea to an eccentric area known as the \ac{PRL} \cite{bronstad_driving_2013}.
This happens gradually over time.
Understanding \ac{PRL} development and the behavioral consequences could potentially help \ac{PLV} improve their oculomotor control in tasks such as reading and visual search~\cite{kwon_rapid_2013}.

Many studies have thus trained sighted participants on a simulated scotoma with the help of the above mentioned gaze-contingent displays, hoping that participants would develop a \ac{PRL}.
However, whereas some studies reported a shift in \ac{PRL} with \ac{SLV} in as fast as three hours~\cite{kwon_rapid_2013,maniglia_method_2020,seitz_we_2020}, others did not \cite{david_effects_2020,copolillo_effects_2017,almutleb_simulation_2018}. 
Longer explicit training times (i.e., 15 to 25 additional hours) has been shown to refine these effects to where oculomotor behavior was comparable to unimpaired controls \cite{kwon_rapid_2013}. 
Maniglia and colleagues~\cite{maniglia_method_2020} conducted a systematic analysis of measures to understand how sighted participants can develop multiple \acp{PRL} and how individual participant re-referencing behavior is not consistent trial-to-trial even when trained. They found that roughly half the participants exhibited saccadic re-referencing even without being instructed to do so~\cite{maniglia_method_2020}.
Kwon and colleagues~\cite{kwon_rapid_2013} showed that explicit training of a scotoma by highlighting estimated \ac{PRL} locations effectively reduced the variance of fixation, much more than when participants were allowed to free-view the stimulus (i.e. the less variance, the more consistent their fixation locations). 

Eye movements under \ac{SLV} can vary drastically depending on how the impairment is presented to the participant~\cite{kwon_rapid_2013,david_effects_2020,chow-wing-bom_worse_2020} and on what task is being studied~\cite{tsank_domain_2017}.
For example, David \emph{et al.}~\cite{david_effects_2020} were able to show that saccade amplitude and fixation duration were significantly larger and longer with a simulated central scotoma (i.e., AMD, $n=32$), whereas the opposite effect was seen for a peripheral scotoma (i.e., RP, $n=32$).
Tsank \emph{et al.}~\cite{tsank_domain_2017} showed that saccade patterns changed for an object-following and a visual-search task, but not when identifying faces.
In another study, McIlreavy \emph{et al.}~\cite{mcilreavy_impact_2012} were able to show that search times for targets and spatial distribution of gaze increased as the size of the simulated scotoma increased, while saccade amplitude and fixation duration remained unaffected.
While both Tsank \emph{et al.}~\cite{tsank_domain_2017} and McIlreavy \emph{et al.}~\cite{mcilreavy_impact_2012} provide insight into the effects of short-term impairment, it remains to be explored to which extent these \ac{SLV} results generalize to real \ac{PLV}.


\subsubsection{XR for Studying Low Vision Participants}

To our surprise, only 20 of the 90 low-vision studies recruited \ac{PLV} (e.g.,~\cite{bowman_individuals_2017,powell_accessibility_2020,miura_virtual_2018,h_hoppe_clevr_2020,lin_evaluating_2014}), all of which were interested in studying how their oculomotor behavior differed from that of sighted people.
Nine of twenty papers were interested in understanding how \ac{VR} could be used to assess behavior of \ac{PLV}.
For example, Bowman and Liu~\cite{bowman_individuals_2017} trained LV participants in a street crossing task. Four out of twelve LV participants were trained with real streets while the other eight were trained with virtual streets using a three-screen \ac{VR} projection system (subtending $168 \times 35$ degrees of visual angle).
Both groups were tested on their street crossing ability in real streets both before and after training.
Before training, all participants demonstrated poor street crossing skills (more than half of the responses were during ``unsafe" times to cross).
After training, over 90 percent of crossing responses were ``safe". 
Training with the \ac{VR} system was comparable to training in real life, demonstrating how \ac{VR} can be a powerful tool for practicing tasks that would otherwise be too dangerous or unfeasible within a laboratory setting \cite{bowman_individuals_2017}.

Lin and colleagues~\cite{lin_evaluating_2014} undertook the VR study with the largest sample size by recruiting 21 participants to perform a reading task while wearing a \ac{VR} \ac{HMD} integrated with \ac{CCTV} magnification software.
The \ac{CCTV} software allowed participants to control the size, brightness, and contrast of the screen.
The study showed that the \ac{HMD} VR-based system led to better reading acuity compared to using a regular screen \cite{lin_evaluating_2014}.
These studies highlight how \ac{XR} can be a tool for training and rehabilitation of skills for \ac{PLV}.

\subsubsection{Common Limitations}

An open question is to which extent low-vision simulations match the visual experience of real \ac{PLV}.
Many \ac{SLV} studies involving sighted people base their simulations on crude approximations of a particular eye condition.
For instance, to simulate a central scotoma, studies would often overlay a (rather salient) gray filled circle over an image that would shift in sync with the participant's saccades.
In contrast, most people with \ac{AMD} are unaware of their scotoma and also have different eye movements from sighted controls because of the scotoma ~\cite{kwon_rapid_2013,seitz_we_2020}.
Recording of eye movements is therefore much more challenging for \ac{PLV} since commercial devices are designed for non-disabled viewing.
Furthermore, \ac{PLV} are often much older than the sighted students typically recruited to participate in \ac{SLV} studies and have more experience using their residual vision for everyday tasks.
It would therefore not be surprising if \ac{PLV} showed differences in eye movement strategies and perceptual learning.
Indeed, the results of previous \ac{SLV} studies with respect to whether participants can learn to develop a \ac{PRL} remains mixed to date (e.g.~\cite{david_effects_2020} vs.~\cite{kwon_rapid_2013}).
In addition, sighted participants recruited for \ac{SLV} typically ranged between 20 and 30 years of age, which is much younger than most people with central vision loss due to \ac{AMD}~\cite{klein_prevalence_2010}.
Perceptual and behavioral differences between \ac{SLV} participants and \ac{PLV} may therefore be partially due to age difference~\cite{yehezkel_crowding_2015}.
Future work could focus on a more direct comparison between the behavioral performance of sighted participants viewing \ac{SLV} with real \ac{PLV}.

A related limitation is the relative lack of \ac{PLV} involvement in this line of research. While 75 of the 90 studies in this category referenced at least one previous study involving \ac{PLV} (Table~\ref{tab:metrics}), we found only one study that grounded their simulation directly in clinical data (e.g.,~\cite{jones_seeing_2020}).
In addition, only a few studies aimed to assess the quality of their simulation by comparing performance to \ac{PLV}.
Future studies could thus work more directly with \ac{PLV} and/or rehabilitation specialists, which may allow for a deeper understanding of how most simulations differ from the daily challenges that \ac{PLV} have to deal with.

\subsection{XR for Augmenting the Residual Vision of People with Low Vision}
\label{sec:XR-VI-augmentation}

Another 33.5\% of papers in our collection focused on the use of \ac{XR} technology to augment and enhance the residual vision of \ac{PLV}.
This can range from handheld or wearables magnifying devices, to applications for smartphones and tablets, to wearable devices like \acp{HMD} and smartglasses.
Whereas \ac{VR} allows for \ac{PLV} to experience otherwise unsafe tasks in a controlled virtual environment, \ac{AR} is better suited as a real-life visual accessibility aid~\cite{gopalakrishnan_use_2020}, as it is allows for real-time interaction with an overlay of the real and digital world (similar to a hearing aid).
Augmentation studies in this category focused on a variety of tasks, ranging from reading to face recognition (e.g.,~\cite{costela_implementation_2021,costela_effect_2021,calabrese_vision_2018}) and obstacle avoidance (e.g.,~\cite{huang_augmented_2019,konik_contribution_2014,angelopoulos_enhanced_2019}). 
Similar to the previous section, most of the studies in this category evaluated their augmentation prototype either directly on \ac{PLV} (e.g.,~\cite{,calabrese_vision_2018,houston_peripheral_2018} or indirectly by using low-vision simulations with sighted participants (e.g.,~\cite{hwang_augmented-reality_2014,van_rheede_improving_2015,foster_safety_2014}).
Some studies also used both (e.g.,~\cite{zhao_understanding_2017}).

\subsubsection{XR for Augmenting Simulated Low Vision}

A small number of studies ($n=17$) used \ac{SLV} not just to study perception and behavior of \ac{PLV}, but to find ways to improve it.
For instance, low-level image manipulations such as increased text magnification and contrast were found to lead to faster reading speeds~\cite{christen_effect_2017}, and enhancing the contours of faces and objects in a visual search task led to faster search times for older participants~\cite{kwon_contour_2012}.
Interestingly, magnification was more beneficial for simulated blurry vision compared to a simulated scotoma~\cite{christen_effect_2017}, whereas contrast enhancement affected reading speed equally across \ac{SLV} conditions.
Similarly, temporal subsampling of an image (``image jitter'') was shown to improve peripheral acuity, word recognition, and facial emotion discrimination~\cite{patrick_temporal_2019,watson_image_2012}.

Many of these studies aimed to understand how smart glasses could be used to help \ac{PLV} (e.g.,~\cite{hwang_augmented-reality_2014,huang_augmented_2019,zhao_understanding_2017}).
See-through \acp{HMD} such as Google Glass and Microsoft Hololens are systems that are commercially available for testing.
Hwang and Peli~\cite{hwang_augmented-reality_2014} measured contrast sensitivity for two conditions with three sighted participants: with or without \ac{AR} edge enhancement and with or without a heavy diffuse film~\cite{hwang_augmented-reality_2014} (Fig.~\ref{fig:xr-lv}A).
The enhancement being tested was the Laplacian edge detection method, where a positive method enhanced edges while a negative method enhanced the surrounding of edges
Contrast sensitivity thresholds had improved with the enhancement method \cite{hwang_augmented-reality_2014}.
Huang and colleagues~\cite{huang_augmented_2019} tested 24 normally sighted participants on a navigation task with a voice based sign reading application for the Hololens.
All participants wore goggles modified with occlusion foils during the task to simulate reduced acuity.
Results indicated that participants walked more slowly and took more time with the sign-reading application.
Overall, participant walked on more direct paths and were more confident with the application.

\subsubsection{XR for Studying Augmentations for Low Vision Participants}

Although results from \ac{SLV} studies are notable, the ultimate goal of an \ac{XR} accessibility aid should be to probably improve the residual vision of real \ac{PLV}.
The majority ($n=64$) of studies in this category thus evaluated their prototypes on appropriate end users.

Zhao and colleagues have published multiple studies on how image enhancement with both \ac{VR} and \ac{AR} can be used to identify and solve challenges that \ac{PLV} face when completing activities of daily living~\cite{zhao_understanding_2017,zhao_effectiveness_2020,zhao_seeingvr_2019,zhao_designing_2019,zhao_designing_2019-1,zhao_cuesee_2016,zhao_foresee_2015}.
``SeeingVR''~\cite{zhao_seeingvr_2019} comprises a set of 14 visual enhancement tools that are designed to make the VR experience of \ac{PLV} more accessible.
The tools include magnification, brightness and contrast controls, edge enhancement, peripheral remapping, text augmentation, depth measuring, text to speech, and more.
When asked to navigate a menu, search for an object, or shoot a moving target while wearing a HTC VIVE, eleven participants completed the tasks much faster and more accurately with SeeingVR than without the application \cite{zhao_seeingvr_2019}.
``ForeSee''~\cite{zhao_foresee_2015,zhao_designing_2019} is an \ac{AR} application that uses a combination of low-level image enhancement methods (e.g., magnification, contrast enhancement, edge enhancement) for reading text in near- and far-distance viewing conditions (Fig.~\ref{fig:xr-lv}B).
``CueSee''~\cite{zhao_cuesee_2016} is an \ac{AR} application to enhance recognition of products on a grocery shelf with the help of five different visual cues including magnification, color enhancement, flashing bounding boxes, and rotation (Fig.~\ref{fig:xr-lv}C).
Participants identified items on a grocery shelf significantly faster using CueSee than without it.

Other groups focused on gaze-contingent image manipulations.
For instance, Calabrese and colleagues~\cite{calabrese_vision_2018} were able to improve face recognition for people with \ac{AMD} via a magnification effect that could highlight and enhance the appearance of a face based on where the user was looking.

Enhancements for mobility, safety, and navigation have been studied using \ac{PLV} wearing modified glasses or \ac{HMD} systems. 
Houston and colleagues~\cite{houston_peripheral_2018} tested the ability of people with \ac{VFL} to navigate a virtual mall while wearing specially designed glasses that could expand the binocular visual field by up to 40 degrees (peripheral prisms).
Twenty-four participants were asked to report obstacles and pedestrians while navigating the virtual mall.
Interestingly, the detection of hazards on the same side of the visual field defect improved significantly for most participants, even without training \cite{houston_peripheral_2018}.
In another study, participants with various diagnosed forms of visual impairment were able to safely complete a stair navigation task with the help of an \ac{AR} \ac{HMD} designed to highlight stair edges~\cite{zhao_designing_2019}.

Whereas most studies focused on enhancing a user's residual vision, others built custom \acp{HMD} to simplify the visual scene.
A notable example is the work by van Rheede and colleagues~\cite{van_rheede_improving_2015}, which built an \ac{HMD} integrated with an infrared depth camera to create a depth map, which was relayed to the user as a grayscale map: the closer the obstacle, the brighter the representation on the display (Fig.~\ref{fig:xr-lv}D).
Participants were then instructed to avoid foam obstacles while navigating a hallway; six of the eleven participants completed the obstacle course without any collisions. 

A number of studies also assessed the usability of the proposed \ac{XR} technologies ($n = 19$).
One study reported that most \ac{PLV} preferred a compact device similar to a regular pair of glasses with buttons for inconspicuous interactions~\cite{mp_hoogsteen_functionality_2020}.
Another study pointed to the portability of a head-mounted system paired with a smartphone as a camera as the preferred form factor for a reading aid~\cite{stearns_design_2018}.
The ForeSee work~\cite{zhao_foresee_2015} highlighted the need to give users the option to choose from several enhancement modes.
Another \ac{AR} study found that alphanumeric representation of information may be better for those with relatively higher acuity, whereas symbolic representation may be better suited for those with worse acuity~\cite{lang_augmented_2020}.
Audio feedback was generally liked by participants as well~\cite{zhao_effectiveness_2020}.

\subsubsection{Common Limitations}

Although most works evaluated their \ac{XR} prototype on \ac{PLV} as a proof of concept, relatively few studies ($n=20$) recorded participant feedback after the study was conducted \ac{PLV} (e.g.,~\cite{zhao_seeingvr_2019,zhao_effectiveness_2020,min_htike_augmented_2021,htike_utilizing_2020}).
However, this may be an important step towards designing more usable accessibility aids that are sensitive to the information needs of \ac{PLV}~\cite{htike_ability_2020}.
For instance, Williams et al.~\cite{williams_just_2014} compared sighted and blind navigation and found that both groups understand navigation differently, leading sighted people to struggle in guiding blind companions.
In addition, people with \ac{BLV} use a combination of devices and technology to complement their existing orientation and mobility skills~\cite{williams_just_2014}, which may lead to a wide variety of navigation styles~\cite{ahmetovic_impact_2019,htike_ability_2020}.
The future \ac{XR} devices being used to help \ac{PLV} with daily tasks sounds promising.
Further collaboration between researchers and end users could benefit device design by augmenting the visual environment based on user-specific needs.

Despite demonstrating an improvement in task performance, many studies reported an increase in trial completion time (e.g.,~\cite{van_rheede_improving_2015,zhao_understanding_2017}), often linked to slower walking speeds or longer search times.
While this may indicate that participants were more careful, it could also indicate increased hesitation or lower confidence when using \ac{VR} and \ac{AR} controls.
In addition, individual differences in visual function (i.e. acuity, thresholds, etc.) could have different effects on performance \cite{lang_augmented_2020}.

\subsection{XR for Studying Perception \& Behavior of People with Prosthetic Vision}
\label{sec:XR-B-behavior}

\Ac{XR} technology has not only been used to augment the vision of \ac{PLV} but also to study the rudimentary vision restored to blind people by a visual neuroprosthesis (``bionic eye''; 15.0\% of papers in our collection)~\cite{fernandez_development_2018}.
Similar to conventional \ac{AR} \acp{HMD}, visual prostheses typically contain an external camera mounted on a pair of glasses that is used to relay the visual scene to the user.
However, in contrast to conventional \ac{AR} \acp{HMD}, visual prostheses also consist of an implantable microstimulator (implanted in the eye or the visual cortex) which decodes the visual information and electrically stimulates neurons in the visual pathway to evoke visual percepts (``phosphenes'').
Existing bionic eyes generally provide an improved ability to localize high-contrast objects and perform basic orientation \& mobility tasks (e.g.,~\cite{stronks_functional_2014}).
While this could be considered a rudimentary form of AR on its own, a good number of studies used \ac{VR} to simulate the perception produced by these devices.

\subsubsection{XR for Simulating Prosthetic Vision}

To investigate functional recovery and experiment with different implant designs, researchers have been developing \ac{XR} prototypes that rely on \ac{SPV} ($n=26$).
The classical method relies on sighted subjects wearing a \ac{VR} headset, who are then deprived of natural viewing and only perceive phosphenes displayed in the \ac{HMD}. This viewing mode has been termed ``transformative reality'' \cite{lui_transformative_2011,lui_transformative_2012} (as opposed to ``altered reality'', which is typically used to describe \ac{SLV} approaches \cite{bao_augmented_2019}).
This allows sighted participants to ``see'' through the eyes of the bionic eye user, taking into account their head and/or eye movements as they explore a virtual environment \cite{kasowski_towards_2021}.

One application of \ac{SPV} is assessing low-level visual function, such as phosphene size~\cite{lu_estimation_2012} and shape~\cite{cao_eye-hand_2017}, by varying stimulus and model parameters.
Stimuli for these tasks are typically presented on a monitor~\cite{lu_estimation_2012}, via \ac{AR} glasses~\cite{caspi_assessing_2015}, or in an \ac{HMD}~\cite{cao_eye-hand_2017}.
In one prominent example~\cite{caspi_assessing_2015}, six sighted volunteers completed a Landolt-C visual acuity task using \ac{SPV}.
Participants wore custom AR glasses to view webcam input that was converted to an $8 \times 8$ pixel image, meant to represent the limited resolution of current retinal implants.
The authors found that well-performing participants developed similar strategies to those employed by real prosthesis users, such as scanning the image using strategic head movements.

The majority of studies focused on slightly more complex tasks such as letter~~\cite{zhao_reading_2011}, word~\cite{fornos_reading_2011}, face~\cite{denis_human_2013,chang_facial_2012}, and object recognition~\cite{zhao_image_2010,wang_cross-task_2018,mace_simulated_2015}. 
This group of studies had the highest average number of subjects ($\mu=21.06\pm12.34)$ when compared to other areas of \ac{SPV} studies. 
In most setups, participants would view \ac{SPV} stimuli in a conventional \ac{VR} \ac{HMD}~($n=8$), but a large portion used a monitor with some sort of motion tracking~($n=7$). 
Surprisingly, although the majority of the tasks used \acp{HMD}, none of the studies allowed for a fully immersive environment that would allow the subject to walk around and interact with it. 
Studies in this category primarily used \ac{SPV} to study basic behavior in these tasks, but some studies also used these tasks to focus on another behavior.
One example of this is the work by Sanchez-Garcia~\emph{et al.}~\cite{sanchez_garcia_influence_2020}. 
In this work, participants ($n=24$) were tasked to find and recognize objects in a scene with different \acp{FOV} ($20^\circ, 40^\circ,$ or $60^\circ $) and number of phosphenes (200 or 500).
The authors showed counter-intuitive results, with a higher \ac{FOV} resulting in significantly worse performance and longer recognition times.
However, they argued that phosphene density may be more important for object recognition than \ac{FOV}, which is consistent with earlier findings~\cite{van_rheede_simulating_2010}.
Another study~\cite{ho_performance_2019} relied on \ac{AR} smart glasses to simulate the artificial vision provided by the PRIMA subretinal implant~\cite{lorach_photovoltaic_2015}. 
This device was developed for people with geographic atrophy as commonly experienced with \ac{AMD}, where vision is first lost in the macula. To simulate this, the authors needed to combine \ac{SPV} in the macula and natural vision in the periphery.
The authors accomplished this by using \ac{AR} smartglasses with black tape occluding the central \ac{FOV} so only the LED overlay was visible in this area.
With this setup, they were able to make testable predictions about the visual acuity to be expected from PRIMA~\cite{lorach_photovoltaic_2015}, which is currently in clinical trials.

Lastly, an increasing number of studies are focusing on spatial cognition tasks, such as obstacle avoidance~\cite{zapf_assistive_2015,zapf_assistive_2016,endo_influence_2019} and wayfinding~\cite{vergnieux_wayfinding_2014}.
By design, these tasks require a more immersive setup that allows for the incorporation of head and eye movements as well as locomotion~\cite{kasowski_immersive_2022}.
The majority of studies identified ($n=5$) incorporated a fully immersive design for their task, while the remaining studies ($n=2$) used \ac{VR} \acp{HMD} but required their subjects to sit/stand in place and use a keyboard/controller to move. 
The studies also used relatively low subject counts ranging from 5 to 17 subjects ($\mu=10.57\pm3.82$). 
The majority of tasks were simply ``proof of concept" experiments showing that users were able to navigate effectively with \ac{SPV}. 
For example, Zapf~\emph{et al.}~\cite{zapf_towards_2015} simulated \ac{RP} by restricting participants to their central $10^\circ$ \ac{FOV} in a virtual environment. 
Participants ($n=11$) completed a variety of tasks consisting of low-lying obstacle circumvention (avoiding traffic cones), static/moving pedestrian avoidance (navigating a corridor with stationary/moving virtual characters), and path following (following a path through parked cars). 
The simulated residual vision was augmented by replacing the area outside of the central $10^\circ$ \ac{FOV} with simulated phosphenes. The authors identified improvements for avoiding low-lying obstacles and following paths, but identified no performance benefits when avoiding stationary head-level targets. 
Surprisingly there was a performance decrease when using \ac{SPV} with moving targets, something the authors attributed to a possible increase in cognitive load. Of the 27 \ac{SPV} studies on behavior, 7 were focused on similar spatial cognition tasks.

\subsubsection{XR for Studying Prosthesis Users}

Rachitskaya~\emph{et al.}~\cite{rachitskaya_computer-assisted_2020} was the only \ac{XR} study to recruit real bionic eye users and to mention consultation with the \ac{BLV} community during development, having utilized an interdisciplinary team that incorporated ophthalmologists and rehabilitation specialists. 
Since there currently is no standardized procedure for vision rehabilitation across different Argus II implantation centers, Rachitskaya and colleagues~\cite{rachitskaya_computer-assisted_2020} developed a Computer-Assisted Rehabilitation Environment (CAREN), which consists of a motion capture system, control software with a \SI{180}{\degree} curved projection screen, a  motion platform, and a treadmill.
Participants ($n=8$) donned a harness, had access to handrails on the treadmill, and were accompanied by a physical therapist. 
After using CAREN twice weekly for 4 weeks, participants showed significant improvements in walking speed and object localization, demonstrating that immersive technology may provide a solution for the standardization of effective rehabilitation approaches to augment bionic eye performance.

\subsubsection{Common Limitations} 

An open question is to which extent prosthetic vision simulations match the visual experience of real prosthesis users.
Similar to \ac{SLV} research, many \ac{SPV} studies base their simulations on crude approximations of prosthetic vision, assuming that each electrode acts as a small independent light source that produces a distinct focal spot of light~\cite{dobelle_artificial_2000}.
However, a growing body of evidence suggests that the vision generated by current visual prostheses is ``fundamentally different'' from natural vision~\cite{erickson-davis_what_2021}, with interactions between implant technology and the neural tissue degrading the quality of the generated prosthetic vision~\cite{fine_pulse_2015,beyeler_model_2019}.
Only a handful of studies ($n=4$ out of 27) have incorporated a great amount of neurophysiological detail into their setup~\cite{josh_psychophysics_2013,vurro_simulation_2014,wang_cross-task_2018,thorn_virtual_2020}, only two of which~\cite{wang_cross-task_2018,thorn_virtual_2020} relied on an established and psychophysically evaluated model of \ac{SPV}.
In addition, the level of immersion offered by most \ac{SPV} studies was relatively low, with many studies simply presenting simulated stimuli on a screen ($n=8$) without taking into account the participant's gaze ($n=17$).
However, most current prostheses provide a very limited \ac{FOV}; for example, the artificial vision generated by Argus II~\cite{luo_argusr_2016}, the most widely adopted retinal implant thus far, is restricted to $10 \times 20$ degrees of visual angle.
This requires users to scan the environment with strategic head movements while trying to piece together the information \cite{erickson-davis_what_2021}.
It is therefore unclear how the findings of most \ac{SPV} studies would translate to real bionic eye users.

Additionally, only a single study in our collection worked with real bionic eye users or rehabilitation specialists~\cite{rachitskaya_computer-assisted_2020}.
\Ac{XR} may offer a unique method for safe training and rehabilitation, but is severely underutilized in comparison to research on \ac{XR} for low vision.


\subsection{XR for Augmenting Prosthetic Vision}
\label{sec:XR-B-augmentation}

Another 15\% of papers in our collection focused on the use of \ac{XR} technology to augment and enhance prosthetic vision, either through simulations ($n=29$) or the use of peripherals and extra sensors to extract visual scene information ($n=5$).

\subsubsection{XR for Augmenting the Visual Scene Using Simulated Prosthetic Vision}

A popular trend for \acf{SPV} is utilizing novel augmentation strategies to aid scene understanding.
One approach is using computer vision to enhance certain image features or regions of interest, at the expense of discarding less important or distracting information.
Various studies have explored strategies based on visual saliency (e.g.,~\cite{parikh_performance_2013}), background subtraction and scene retargeting (e.g.,~\cite{li_optimized_2018}), and depth mapping to highlight nearby obstacles(e.g.,~\cite{lieby_substituting_2011,mccarthy_mobility_2015,kartha_prosthetic_2020}).
For instance, McCarthy and colleagues~\cite{mccarthy_mobility_2015} used an RGB-D camera mounted on a pair of \ac{AR} glasses to augment the visual scene with depth information (Fig.~\ref{fig:spv}B).
The study used a custom augmented reality setup utilizing an \ac{HMD} with an attached stereo camera. The images from the camera were sent to a laptop on the participant's back, and were processed into a simplified model of bionic vision using a pixel display with twenty phosphenes.
Among their tested image processing strategies, augmented depth proved the most effective at highlighting hazards in the path; this mode modified the depth information from the stereo cameras to detect objects while simultaneously removing the ground from the scene.
They found a significantly reduced rate of collisions, even in the presence of low-contrast trip hazards.
These findings were later evaluated with real bionic eye users~\cite{barnes_enhancing_2015}.

The majority ($n=18$) of \ac{SPV} studies in this category used monitors~($n=3$), VR headsets~($n=14$), and \ac{AR} glasses~($n=1$) to improve performance on recognition tasks, such as identifying faces~\cite{chang_facial_2012, wang_face_2014,irons_face_2017}, text~\cite{denis_simulated_2014, paraskevoudi_full_2021}, and objects~\cite{li_optimized_2018, wang_image_2016}.
For instance, a number of studies \cite{chang_facial_2012, wang_face_2014,irons_face_2017} highlighted through simulations that face caricaturing, where prominent facial features are highlighted or enhanced, can improve face recognition for sighted subjects viewing \ac{SPV}.
Studies focused on recognition applied various enhancements including contrast enhancement~\cite{chang_effect_2010}, saliency algorithms~\cite{li_image_2018,wang_image_2016}, edge/foreground extraction~\cite{han_object_2015,lui_transformative_2011}, and facial landmark extraction~\cite{bollen_simulating_2019}.

8 \ac{SPV} studies focused on spatial cognition tasks, including wayfinding~\cite{vergnieux_simplification_2017, van_rheede_simulating_2010}, obstacle avoidance~\cite{mccarthy_mobility_2015}, and environmental search~\cite{parikh_performance_2013}.
For instance, van Rheede and colleagues~\cite{van_rheede_simulating_2010} used gaze-contingent \ac{SPV} to measure acuity, object recognition, and mobility (Fig.~\ref{fig:spv}A).
They found that using a region-of-interest view improved acuity while a wide \ac{FOV} was better for mobility, highlighting the use of testing multiple forms of enhancement with various tasks \cite{van_rheede_simulating_2010}.
In a similar manner to the spatial behavioral \ac{SPV} studies, the majority ($n=6$) of augmented \ac{SPV} studies also used \ac{VR} \acp{HMD} with two studies using \ac{AR} smartglasses~\cite{weiland_smart_2012,mccarthy_mobility_2015,parikh_performance_2013}.Most of the studies were fully immersive ($n=6$), but two used \ac{VR} \acp{HMD} without positional tracking~\cite{vergnieux_simplification_2017,vergnieux_wayfinding_2014}. Out of these 8 studies, only 1 study used eye-tracking~\cite{van_rheede_simulating_2010} or presented monocular stimuli~\cite{parikh_performance_2013}. These studies also suffered from relatively low subject counts ranging from 4 to 19 subjects ($\mu=10.75\pm4.62$). 

The remaining 3 studies identified simulation studies focused on low-level visual functioning ~\cite{bermudez-cameo_rgb-d_2017,al-atabany_extraspectral_2018, titchener_gaze_2018}.
Two of these studies used \ac{AR} for enhancing an \ac{SPV} scene with one using infrared (IR) overlays for counting people/actions in a scene,~\cite{al-atabany_extraspectral_2018}, and the other using RGB-D cameras for depth overlays in a target localization task~\cite{bermudez-cameo_rgb-d_2017}.
The third study used a \ac{VR} \ac{HMD} with eye tracking to study the effects of gaze in a simulated retinal prosthesis~\cite{titchener_gaze_2018}.
All three studies performed basic low-level visual function testing but suggested the augmentations could be valuable for more advanced scene understanding.

\subsubsection{XR for Studying Augmentations for Prosthetic Vision}

While a bionic eye is technically itself a technology that augments vision, several studies focused on augmentation strategies outside the basic stimulation patterns of the device. 
This includes thermal imaging~\cite{zagar_low_2010,he_trade-off_2019}, audio-visual cross-modal mapping\cite{stiles_multisensory_2021}, and depth detection with object segmentation~\cite{kartha_prosthetic_2020}.
For example, Sadeghi \emph{et al.}~\cite{sadeghi_glow_2021} tested the ability of bionic eye users to perform a series of practical tasks (e.g., identifying hot objects, estimating the distance to a nearby person) while using a thermal camera.
The study highlighted improved performance across all tested tasks, including tasks where thermal integration would be considered an obvious benefit (e.g., identifying the closer side of a hot cup, identifying a missing bowl that was heated), but also tasks such as determining whether an escalator was moving towards or away from them.
In another study by Kartha \emph{et al.}~\cite{kartha_prosthetic_2020}, Argus II users completed various tasks with a distance-filtered input.
In this study, the removal of distant clutter was able to improve participant performance across a variety of tasks including size, depth, and walking direction discrimination. While these augmentations were relatively simple, they were still able to improve performance across a variety of tasks.
These results are promising, and present the possibility for more advanced augmentation methods to be useful in the future.

\subsubsection{Common Limitations}

Similar to the previous section, \ac{SPV} studies used to augment prosthetic vision relied on the same crude approximations of the visual experience of real bionic eye users.
Although there is no shortage of publications that demonstrate a proof-of-concept augmentation strategy, only a few studies discussed the usability aspects of their proposed technology (e.g.,~\cite{ sadeghi_glow_2021, kartha_prosthetic_2020}).
Additionally, very few \ac{SPV} studies allowed participants to move around in an immersive way ($n=6$ out of 29), and all those that did were focused on spatial cognition.
Typical real-life scenarios cannot be mastered while stationary, and future studies may benefit from allowing participants to move around their environment.
Many studies used \ac{SPV} to assess the benefit of their proposed technology, but only two used models based on neuroscience, considered gaze ($n=6$), or presented monocular stimuli ($n=5$).

Because the involvement of real bionic eye users remains limited (500 implantees worldwide) and challenging (e.g., constant assistance, increased setup time, travel cost), it is not surprising that most behavioral results were reported on a relatively small sample size limited to one to five participants.
While \ac{XR} technology in combination with \ac{SPV} may provide a more cost-effective alternative to prototyping novel augmentation strategies \cite{kasowski_towards_2021}, future studies should consider a more direct comparison between their theoretical predictions and the visual experience reported by real bionic eye users \cite{erickson-davis_what_2021}.


\section{Discussion}

\subsection{\emph{The Main Types of XR Technologies Used in BLV Research}}

\ac{VR} and \ac{AR} \ac{HMD}s were the most common ($n = 58$ and $n = 55$, respectively). 
\ac{AR} \ac{HMD}s have a slight advantage over \ac{VR} given that they can be used in real-life situations, rather than simulations. \ac{AR} can also be used as an accessibility tool, such as enhancing text (e.g.,~\cite{huang_augmented_2019,zhao_recognition_2017}; similar to \ac{VR}), or, more notably, highlighting obstacles while navigating a real environment (e.g.,~\cite{hicks_depth-based_2013,van_rheede_improving_2015}). 

\Ac{VR} wearables were by far the most popular device type among the studies in our corpus, prevalent in both low vision and prosthetic vision research (Table~\ref{tab:metrics}).
Interest in these devices has been more or less constant over the last decade, but has seen a recent increase in 2020--21 (Table~\ref{tab:year-devices}).
\Ac{VR} devices have the advantage of allowing researchers to fully control the visual stimuli presented to the participants, which can make for a flexible testbed for prototypes of near-future visual accessibility aids~\cite{h_hoppe_clevr_2020, zhao_seeingvr_2019}.
They also offer a safe method for testing behavior that would otherwise be too dangerous for the participant, such as crossing streets~\cite{bowman_individuals_2017, thevin_x-road_2020,rachitskaya_computer-assisted_2020} or driving with low vision~\cite{alberti_visual_2014}.
In addition, \ac{XR} technologies afford the ability to simulate prosthetic vision without the need for invasive surgical procedures~\cite{xia_adaptation_2015, sanchez_garcia_influence_2020,thorn_virtual_2020}.
The most commonly used \ac{VR} \acp{HMD} included the HTC Vive, Fove 0, and Oculus Rift (e.g.~\cite{h_hoppe_clevr_2020,zhao_seeingvr_2019,jones_seeing_2020,chow-wing-bom_worse_2020,kvansakul_sensory_2020}).

Desktop monitors were another trusted device type with constant interest over the years (Table~\ref{tab:year-devices}). Monitors can be particularly useful if they are used as a gaze-contingent display to study changes in eye movements (Table~\ref{tab:metrics}).
Nonelectronic wearables (glasses, goggles, etc.) are the inexpensive, but third most common option in our dataset (e.g.,~\cite{wood_effect_2010,copolillo_effects_2017,alberti_binocular_2018}).

\subsection{\emph{The Experimental Tasks Studied with XR}}

For both low and prosthetic vision, visual search and recognition tasks were the most common ($n=114$), followed by spatial cognition  tasks related to orientation \& mobility ($n=78$) and low-level visual function testing ($n=35$).
While acuity is known as the standard for assessing visual function, the degree to which it associates to other tasks still remains to be explored, especially for changes in acuity with \ac{SLV} and \ac{SPV}.
Future work would benefit from comparing performance across tasks not limited to low-level visual function. Future development of  \ac{AR} \acp{HMD} could potentially improve \ac{BLV} function for a range of tasks rather than one. 

With the exception of low vision augmentation studies, it is worth noting that most work involved sighted participants viewing \ac{SLV} ($n=95$ out of 166) or \ac{SPV} ($n=55$ out of 61).
While this may reflect difficulty recruiting \ac{PLV}, simulations have so far proved a valuable tool to enable large-scale behavioral studies and the quick prototyping of novel augmentation strategies.


\subsection{\emph{Key Challenges and Scientific Gaps}}

\ac{XR} technologies have seen major improvements in functionality and costs over the past decade, and the interest to use these devices in \ac{BLV} research has risen accordingly.

However, there are a number of challenges and limitations that were common across the different research areas:

\begin{itemize}[topsep=0pt]
    \item It is unclear to what extent simulations of low vision and prosthetic vision match the visual experience of people with \ac{BLV}.
    Only a few studies thoroughly grounded their simulations in real patient data (e.g.,~\cite{jones_seeing_2020,thorn_virtual_2020}, and studies using simulations had much younger participants than the target \ac{BLV} group ($\mu=29.2$ and $\mu=59.72$ years, respectively).
    Many studies used crude approximations of the underlying eye condition and ignored the immersiveness of their simulation. For example, only 40\% of studies used a gaze-contingent display, meaning that sighted participants could artificially increase their \ac{FOV} through eye movements. Ignoring this aspect could result in simulated participants performing much better on tasks than those with the condition being studied. Addressing this could lead to more insightful simulations. 

    \item While many studies ($n=76$) used \ac{BLV} participants to test the performance of a new system, very few studies consulted with the \ac{BLV} community ($n=46$) during the early phases of their study. Instead, the majority of studies focused on technical developments such as exploring different computer vision and enhancement techniques (e.g.,~\cite{hommaru_walking_2020,van_rheede_improving_2015,mccarthy_mobility_2015}) and reporting quantitative measures such as mobility efficiency and errors, obstacle detection rates, and clinical visual measurements \cite{houston_peripheral_2018,hicks_depth-based_2013,barhorst-cates_let_2017}. Less emphasis has been placed on understanding the usability and suitability of these aids in people with different levels of residual vision and underlying conditions, or whether or not these accessibility aids address the information needs of \ac{BLV} users. 

    \item Even fewer studies~($n=31$) collected \ac{BLV} participants' opinions after the study. While the proposed systems may have improved performance in specific tasks, the systems must also be user-friendly and avoid steep learning curves. 
    Struggling to adapt to new technologies may limit device use or prevent end users from acquiring the necessary skill set to fully utilize a new accessibility aid.
    Surveying user preferences at the end of a performance evaluation could lead to insights that may increase usability and adoption rates.
\end{itemize}

Unfortunately, to the best of our knowledge, none of the reviewed devices and applications have found widespread adoption.
While the underlying reasons may be nuanced and manifold,
many studies in our collection seem poised for success in the near future (e.g.~\cite{hwang_augmented-reality_2014,mccarthy_mobility_2015,zhao_seeingvr_2019}).
We are hopeful that addressing the highlighted gaps in the existing literature will lead to increased usability of different XR-based accessibility aids.


\section{Conclusion}

In conclusion, our systematic review has highlighted the benefits of \ac{XR} technology for \ac{BLV} research, but challenges still remain.
By broadening end-user participation to early stages of the design process and shifting the focus from behavioral performance to qualitative assessments of usability, future research has the potential to develop \ac{XR} technologies that may not only allow for studying vision loss, but also enable novel visual accessibility aids with the potential to impact the lives of millions of people living with vision loss. 


\bibliographystyle{unsrt}
\bibliography{2022-XR-Review-Final,2022-XR-Review-Additional-Refs.bib}

\newpage

\end{document}